\documentclass[runningheads]{llncs}
\usepackage[T1]{fontenc}
\usepackage{graphicx}
\usepackage{amssymb}
\usepackage{amsmath}
\usepackage{mathtools}
\usepackage{bbm}
\usepackage{enumitem}
\usepackage{tabularx}
\usepackage{booktabs}
\usepackage{multirow}
\usepackage{amsfonts}
\usepackage{makecell}
\usepackage{bm}
\usepackage{cite}
\usepackage{array}
\usepackage{url}
\usepackage[misc]{ifsym}
\usepackage{bbding}

\usepackage[colorlinks,
linkcolor=blue,
anchorcolor=blue,
citecolor=blue]{hyperref}

\usepackage{graphicx}

\begin{document}
\title{SimPLe: Similarity-Aware Propagation Learning for Weakly-Supervised Breast Cancer Segmentation in DCE-MRI}
\titlerunning{SimPLe for Weakly-Supervised Segmentation}
%
\author{Yuming Zhong \and Yi Wang\Envelope}
%

\authorrunning{Y. Zhong et al.}
%
\institute{
	Smart Medical Imaging, Learning and Engineering (SMILE) Lab,\\
	Medical UltraSound Image Computing (MUSIC) Lab,\\
	School of Biomedical Engineering,
	Shenzhen University Medical School,\\
	Shenzhen University, Shenzhen, China\\
	\email{onewang@szu.edu.cn} \\
}
\maketitle              
\begin{abstract}
Breast dynamic contrast-enhanced magnetic resonance imaging (DCE-MRI) plays an important role in the screening and prognosis assessment of high-risk breast cancer.
The segmentation of cancerous regions is essential useful for the subsequent analysis of breast MRI.
To alleviate the annotation effort to train the segmentation networks, we propose a weakly-supervised strategy using extreme points as annotations for breast cancer segmentation.
Without using any bells and whistles, our strategy focuses on fully exploiting the learning capability of the routine training procedure, i.e., the \textit{train} - \textit{fine-tune} - \textit{retrain} process.
The network first utilizes the pseudo-masks generated using the extreme points to \textit{train} itself, by minimizing a contrastive loss, which encourages the network to learn more representative features for cancerous voxels.
Then the trained network \textit{fine-tune}s itself by using a similarity-aware propagation learning (SimPLe) strategy, which leverages feature similarity between unlabeled and positive voxels to propagate labels.
Finally the network \textit{retrain}s itself by employing the pseudo-masks generated using previous fine-tuned network.
The proposed method is evaluated on our collected DCE-MRI dataset containing 206 patients with biopsy-proven breast cancers.
Experimental results demonstrate our method effectively fine-tunes the network by using the SimPLe strategy, and achieves a mean Dice value of 81\%.
\textit{Our code is publicly available at} \url{https://github.com/Abner228/SmileCode}.

\end{abstract}

\section{Introduction}
\begin{figure}[t]
	\centering
	\includegraphics[width=1.0\textwidth]{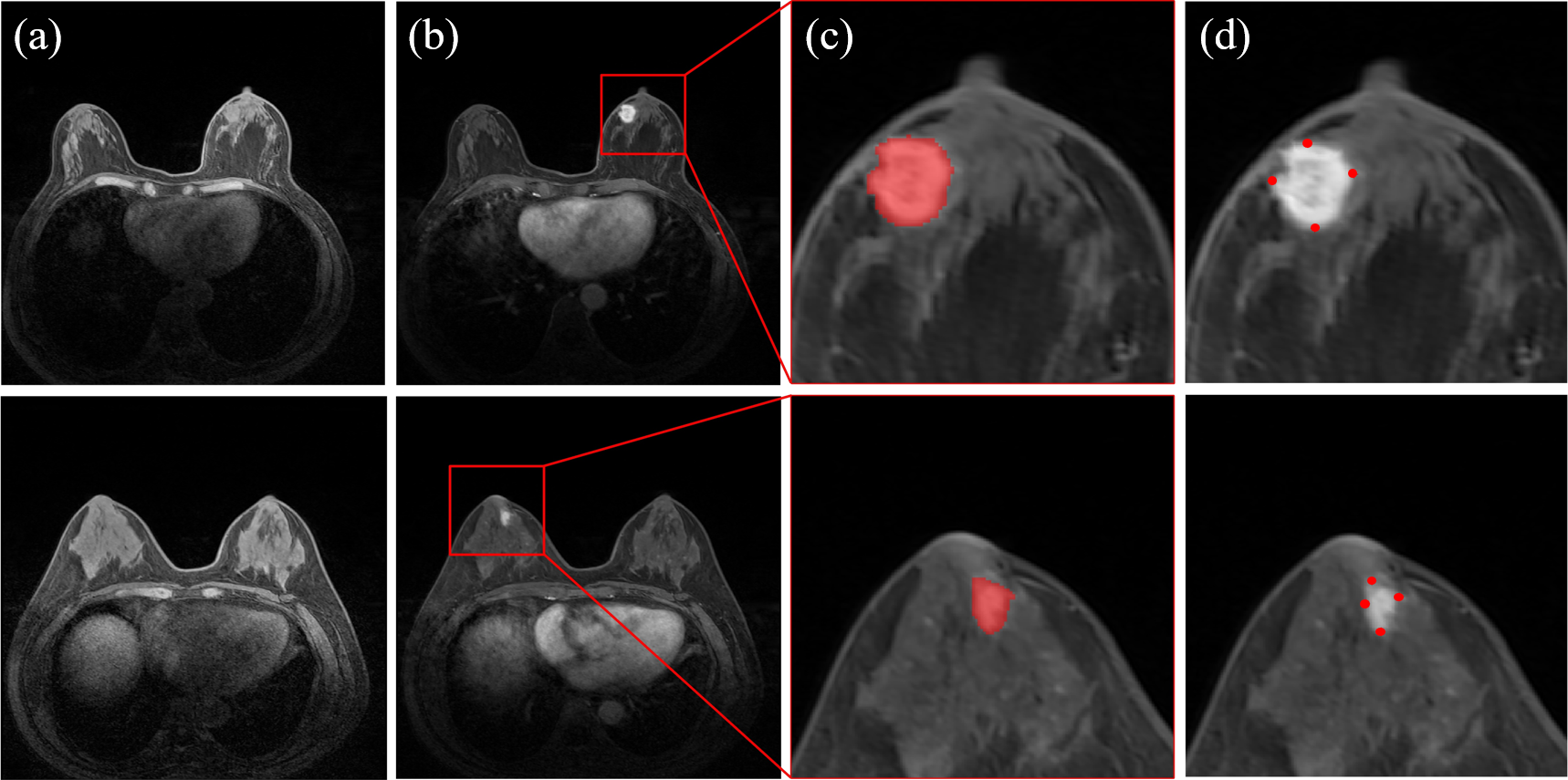}
	\caption{Breast MRI and different annotations: (a) T1-weighted images, (b) corresponding contrast-enhanced images, (c) the cancer annotation with full segmentation masks, and (d) the cancer annotation using extreme points (\textit{note that to facilitate the visualization, here we show the extreme points in 2D images, our method is based on 3D}). }
	\label{fig:intro}
\end{figure}

Breast cancer is the most common cause of cancer-related deaths among women all around the world~\cite{giaquinto2022breast}.
Early diagnosis and treatment is beneficial to improve the survival rate and prognosis of breast cancer patients.
Mammography, ultrasonography, and magnetic resonance imaging (MRI) are routine imaging modalities for breast examinations~\cite{LEE201018}.
Recent clinical studies have proven that dynamic contrast-enhanced (DCE)-MRI has the capability to reflect tumor morphology, texture, and kinetic heterogeneity~\cite{kim2020kinetic}, and is with the highest sensitivity for breast cancer screening and diagnosis among current clinical imaging modalities~\cite{mann2019breast}.
The basis for DCE-MRI is a dynamic T1-weighted contrast enhanced sequence (Fig.~\ref{fig:intro}).
T1-weighted acquisition depicts enhancing abnormalities after contrast material administration, that is, the cancer screening is performed by using the post-contrast images.
Radiologists will analyze features such as texture, morphology, and then make the treatment plan or prognosis assessment.
Computer-aided feature quantification and diagnosis algorithms have recently been exploited to facilitate radiologists analyze breast DCE-MRI~\cite{jiang2021artificial, sheth2020artificial}, in which automatic cancer segmentation is the very first and important step.

To better support the radiologists with breast cancer diagnosis, various segmentation algorithms have been developed~\cite{rezaei2021review}.
Early studies focused on image processing based approaches by conducting graph-cut segmentation~\cite{zheng2007segmentation} or analyzing low-level hand-crafted features~\cite{ashraf2012multi, gubern2015automated, MILITELLO2022103113}.
These methods may encounter the issue of high computational complexity when analyzing volumetric data, and most of them require manual interactions.
Recently, deep-learning-based methods have been applied to analyze breast MRI.
Zhang \textit{et al}.~\cite{zhang2018hierarchical} proposed a mask-guided hierarchical learning framework for breast tumor segmentation via  convolutional neural networks (CNNs), in which breast masks were also required to train one of CNNs.
This framework achieved a mean Dice value of 72\%.
Li \textit{et al}.~\cite{li2019learning} developed a multi-stream fusion mechanism to analyze T1/T2-weighted sequences, and obtained a Dice result of 77\%.
Gao \textit{et al}.~\cite{gao2019dense} proposed a 2D CNN architecture with designed attention modules, and got a Dice result of 81\%.
Zhou \textit{et al}.~\cite{zhou2022three} employed a 3D affinity learning based multi-branch ensemble network for the segmentation refinement and generated 78\% Dice value.
Wang \textit{et al}.~\cite{WANG2021102607} integrated a combined 2D and 3D convolution module and a contextual pyramid into U-net to obtain a Dice result of 76\%.
Wang \textit{et al}.~\cite{wang2021breast} proposed a tumor-sensitive synthesis module to reduce false segmentation and obtained 78\% Dice value.
To reduce the huge annotation burden for the segmentation task, Zeng \textit{et al}.~\cite{zeng2022reciprocal} presented a semi-supervised strategy to segment the manually cropped DCE-MRI images, and attained a Dice value of 78\%.

Although \cite{zeng2022reciprocal} has been proposed to alleviate the annotation effort, to acquire the voxel-level segmentation masks is still time-consuming and laborious, see Fig.~\ref{fig:intro}(c).
Weakly-supervised learning strategies such as extreme points~\cite{roth2021going, dorent2021inter}, bounding box~\cite{du2023weakly} and scribbles~\cite{dorent2020scribble} can be promising solutions.
Roth \textit{et al}.~\cite{roth2021going} utilized extreme points to generate scribbles to supervise the training of the segmentation network.
Based on~\cite{roth2021going}, Dorent \textit{et al}.~\cite{dorent2021inter} introduced a regularized loss~\cite{dorent2020scribble} derived from a Conditional Random Field (CRF) formulation to encourage the prediction consistency over homogeneous regions.
Du \textit{et al}.~\cite{du2023weakly} employed bounding boxes to train the segmentation network for organs.
However, the geometric prior used in~\cite{du2023weakly} can not be an appropriate strategy for the segmentation of lesions with various shapes.
To our knowledge, currently only one weakly-supervised work~\cite{meng2022volume} has been proposed for breast mass segmentation in DCE-MRI.
This method employed three partial annotation methods including single-slice, orthogonal-slice (i.e., 3 slices) and interval-slice ($\sim$6 slices) to alleviate the annotation cost, and then constrained segmentation by estimated volume using the partial annotation.
The method obtained a Dice value of 83\% using the interval-slice annotation, on a testing dataset containing only 28 patients.

In this study, we propose a simple yet effective weakly-supervised strategy, by using extreme points as annotations (see Fig.~\ref{fig:intro}(d)) to segment breast cancer.
Specifically, we attempt to optimize the segmentation network via the conventional \textit{train} - \textit{fine-tune} - \textit{retrain} process.
The initial training is supervised by a contrastive loss to pull close positive voxels in feature space.
The fine-tune is conducted by using a similarity-aware propagation learning (SimPLe) strategy to update the pseudo-masks for the subsequent retrain.
We evaluate our method on a collected DCE-MRI dataset containing 206 subjects.
Experimental results show our method achieves competitive performance compared with fully supervision, demonstrating the efficacy of the proposed SimPLe strategy.

\section{Method}
\begin{figure}[t]
	\centering
	\includegraphics[width=1.0\textwidth]{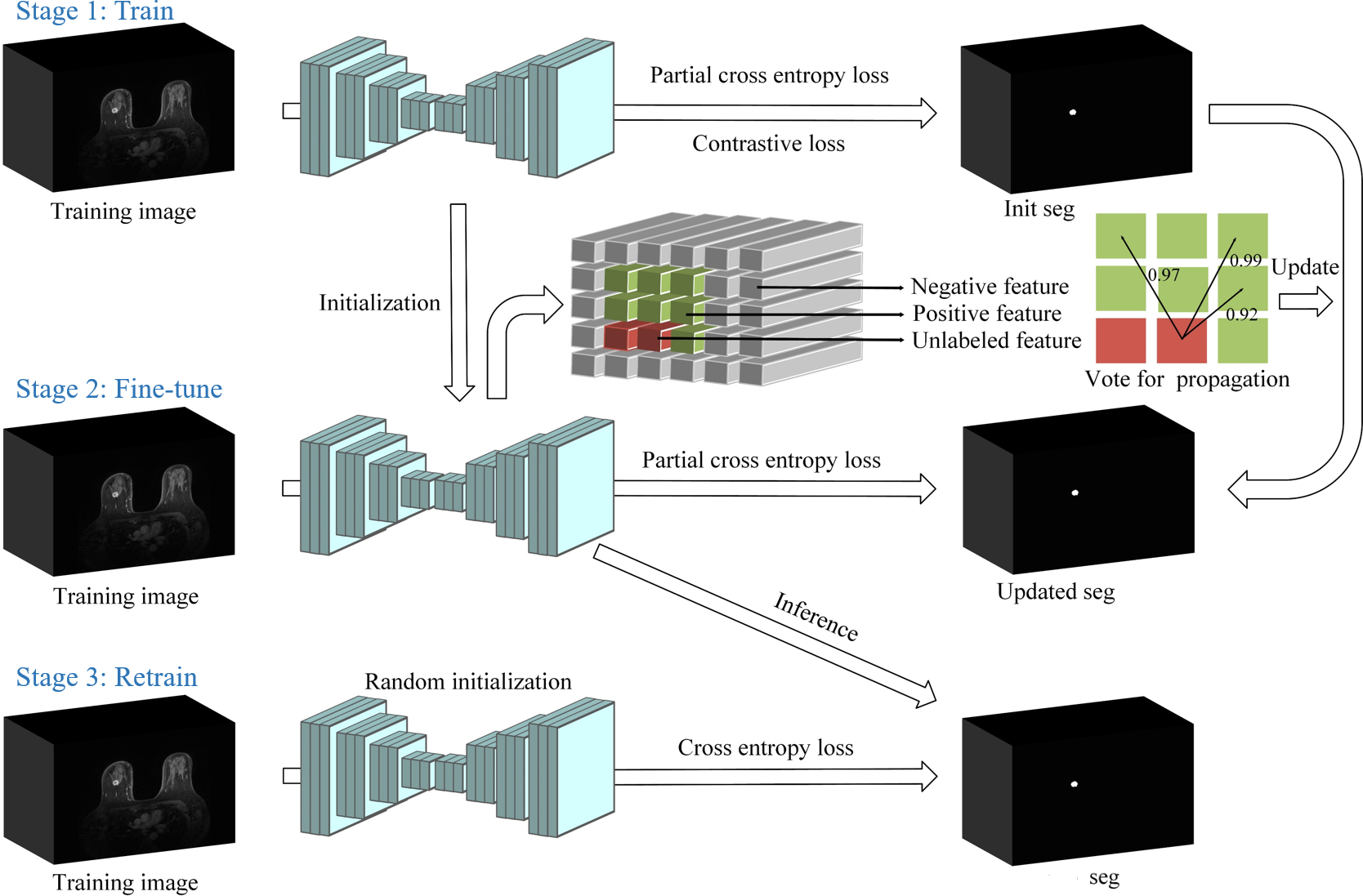}
	\caption{The schematic illustration of the proposed similarity-aware propagation learning (SimPLe) and the \textit{train} - \textit{fine-tune} - \textit{retrain} procedure for the breast cancer segmentation in DCE-MRI.}
	\label{fig2}
\end{figure}

The proposed SimPLe strategy and the \textit{train} - \textit{fine-tune} - \textit{retrain} procedure is illustrated in Fig.~\ref{fig2}.
The extreme points are defined as the left-, right-, anterior-, posterior-, inferior-, and superior-most points of the cancerous region in 3D.
The initial pseudo-masks are generated according to the extreme points by using the random walker algorithm.
The segmentation network is firstly trained based on the initial pseudo-masks.
Then SimPLe is employed to fine-tune the network and update the pseudo-masks.
At last, the network is retrained from random initialization using the updated pseudo-masks.

\subsection{Generate Initial Pseudo-masks}
We use the extreme points to generate pseudo-masks based on random walker algorithm~\cite{grady2006random}.
To improve the performance of random walker, according to~\cite{roth2021going}, we first generate scribbles by searching the shortest path on gradient magnitude map between each extreme point pair via the Dijkstra algorithm~\cite{dijkstra1959note}.
After generating the scribbles, we propose to dilate them to increase foreground seeds for random walker.
Voxels outside the bounding box (note that once we have the six extreme points, we have the 3D bounding box of the cancer) are expected to be the background seeds.
Next, the random walker algorithm is used to produce a foreground probability map $\widehat{Y}: \Omega \subset \mathbb{R}^{3} \rightarrow [0,1]$, where $\Omega$ is the spatial domain.
To further increase the area of foreground, the voxel at location $\vec{k}$ is considered as new foreground seed if $\widehat{Y}(\vec{k})$ is greater than 0.8 and new background seed if $\widehat{Y}(\vec{k})$ is less than 0.1.
Then we run the random walker algorithm repeatedly.
After seven times iterations, we set foreground in the same way via the last output probability map.
Voxels outside the bounding box are considered as background.
The rest of voxels remain unlabeled.
This is the way initial pseudo-masks $Y_{init}: \Omega \subset \mathbb{R}^{3} \rightarrow \{0,1,2\}$ generated,
where 0, 1 and 2 represent negative, positive and unlabeled.

\subsection{Train Network with Initial Pseudo-masks}
Let $X: \Omega \subset \mathbb{R}^{3} \rightarrow \mathbb{R}$ denotes a training volume. 
Let $f$ and $\theta$ be network and its parameters, respectively.
A simple training approach is to minimize the partial cross entropy loss $\mathcal{L}_{pce}$, which is formulated as:
\begin{equation}
\mathcal{L}_{pce} = 
- \sum_{\mathclap{Y_{init}(\vec{k})=0}}log(1-f(X; \theta)(\vec{k}))
- \sum_{\mathclap{Y_{init}(\vec{k})=1}}log(f(X; \theta)(\vec{k})).
\end{equation}
Moreover, supervised contrastive learning is employed to encourage voxels of the same label to gather around in feature space.
It ensures the network to learn discriminative features for each category.
Specifically, features corresponding to $N$ negative voxels and $N$ positive voxels are randomly sampled, then the contrastive loss $\mathcal{L}_{ctr}$ is minimized:
\begin{equation}
\begin{split}
\mathcal{L}_{ctr} = 
-\frac{1}{2N-1}
	&\sum_{\mathclap{\vec{k_n} \in \mathcal{N}(\vec{k})}}
	log(1-\sigma(sim(
	\mathcal{Z}(\vec{k}), \mathcal{Z}(\vec{k_n})) /\tau
	))\\
	&-\frac{1}{2N-1} \sum_{\mathclap{\vec{k_p} \in \mathcal{P}(\vec{k})}}
	log(\sigma(sim(
	\mathcal{Z}(\vec{k}), \mathcal{Z}(\vec{k_p})) /\tau
	)),
\end{split}
\end{equation}
where $\mathcal{P}(\vec{k})$ denotes the set of points with the same label as the voxel $\vec{k}$ and $\mathcal{N}(\vec{k})$ denotes the set of points with the different label.
$\mathcal{Z}(\vec{k})$ denotes the feature vector of the voxel at location $\vec{k}$.
$sim(\cdot,\cdot)$ is the cosine similarity function.
$\sigma$ denotes sigmoid function.
$\tau$ is a temperature parameter.

To summarize, we employ the sum of the partial cross entropy loss $\mathcal{L}_{pce}$ and the contrastive loss $\mathcal{L}_{ctr}$ to train the network with initial pseudo-masks:
\begin{equation}
\mathcal{L}_{train} = \mathcal{L}_{pce} + \mathcal{L}_{ctr}.
\end{equation}

\subsection{SimPLe-based Fine-tune and Retrain}
The performance of the network trained by the incomplete initial pseudo-masks is still limited.
We propose to fine-tune the entire network using the pre-trained weights as initialization.
The fine-tune follows the SimPLe strategy which evaluates the similarity between unlabeled voxels and positive voxels to propagate labels to unlabeled voxels.
Specifically, $N$ positive voxels are randomly sampled as the referring voxel.
For each unlabeled voxel $\vec{k}$,
we evaluate its similarity with all referring voxels:
\begin{equation}
\mathcal{S}(\vec{k}) = \sum_{i=1}^{N}{\mathbb{I}\{
sim(\mathcal{Z}_{\vec{k}}, \mathcal{Z}_{i}) > \lambda \} },
\end{equation}
where $\mathbb{I}(\cdot)$ is the indicator function, which is equal to 1 if the cosine similarity is greater than $\lambda$ and 0 if less.
If $\mathcal{S}(\vec{k})$ is greater than $\alpha N$, the voxel at location $\vec{k}$ is considered as positive.
Then the network is fine-tuned using the partial cross entropy loss same as in the initial train stage.
The loss function $\mathcal{L}_{finetune}$ is formulated as:
\begin{equation}
\mathcal{L}_{finetune} = \mathcal{L}_{pce}
- w \cdot \sum_{\mathclap{\mathcal{S}(\vec{k}) > \alpha N}}
log(f(X; \theta)(\vec{k})),
\end{equation}
where $w$ is the weighting coefficient that controls the influence of the pseudo labels.
To reduce the influence of possible incorrect label propagation, pseudo labels for unlabeled voxels are valid only for the current iteration when they are generated.

After the fine-tune completed, the network generates binary pseudo-masks for every training data, which are expected to be similar to the ground-truths provided by radiologists.
Finally the network is retrained from random initialization by minimizing the cross entropy loss with the binary pseudo-masks.

\section{Experiments}
\subsubsection{Dataset.}
We evaluated our method on an in-house breast DCE-MRI dataset collected from the Cancer Center of Sun Yat-Sen University.
In total, we collected 206 DCE-MRI scans with biopsy-proven breast cancers.
All MRI scans were examined with 1.5T MRI scanner.
The DCE-MRI sequences (TR/TE = 4.43ms/1.50ms, and flip angle=10$^{\circ}$) using gadolinium-based contrast agent were performed with the T1-weighted gradient echo technique, and injected 0.2ml/kg intravenously at 2.0ml/s followed by 20ml saline.
The DCE-MRI volumes have two kinds of resolution, 0.379$\times$0.379$\times$1.700 $\text{mm}^3$ and 0.511$\times$0.511$\times$1.000 $\text{mm}^3$.

All cancerous regions and extreme points were manually annotated by an experienced radiologist via ITK-SNAP~\cite{py06nimg} and further confirmed by another radiologist.
We randomly divided the dataset into 21 scans for training and the remaining scans for testing.
Before training, we resampled all volumes into the same target spacing 0.600$\times$0.600$\times$1.000 $\text{mm}^3$
and normalized all volumes as zero mean and unit variance.

\subsubsection{Implementation Details.}
The framework was implemented in PyTorch, using a NVIDIA GeForce GTX 1080 Ti with 11GB of memory.
We employed 3D U-net~\cite{cciccek20163d} as our network backbone.\\
$\bullet$
\textit{Train}: The network was trained by stochastic gradient descent (SGD) for 200 epochs, with an initial learning rate $\eta=0.01$.
The ploy learning policy was used to adjust the learning rate, $(1 - \text{epoch} / 200)^{0.9}$.
The batch size was 2, consisting of a random foreground patch and a random background patch located via initial segmentation $Y_{init}$.
Such setting can help alleviate class imbalance issue.
The patch size was $128 \times 128 \times 96$.
For the contrastive loss, we set $\text{N} = 100$, temperature parameter $\tau = 0.1$.\\
$\bullet$
\textit{Fine-tune}:
We initialized the network with the trained weights.
We trained it by SGD for 100 iterations, with $\eta=0.0001$.
The ploy learning policy was also used.
For the SimPLe strategy, we set $N = 100, \lambda = 0.96, \alpha = 0.96, w = 0.1$.
After fine-tuned, the last weights were used to generate the binary pseudo-masks.\\
$\bullet$
\textit{Retrain}: The training strategy was the same as the initial train stage.\\
$\bullet$
\textit{Inference}:
A sliding window approach was used.
The window size equaled the patch size used during training. 
Stride was set as half of a patch.

\subsubsection{Quantitative and Qualitative Analysis.}

\begin{figure}[t]
	\centering
	\includegraphics[width=1.0\textwidth]{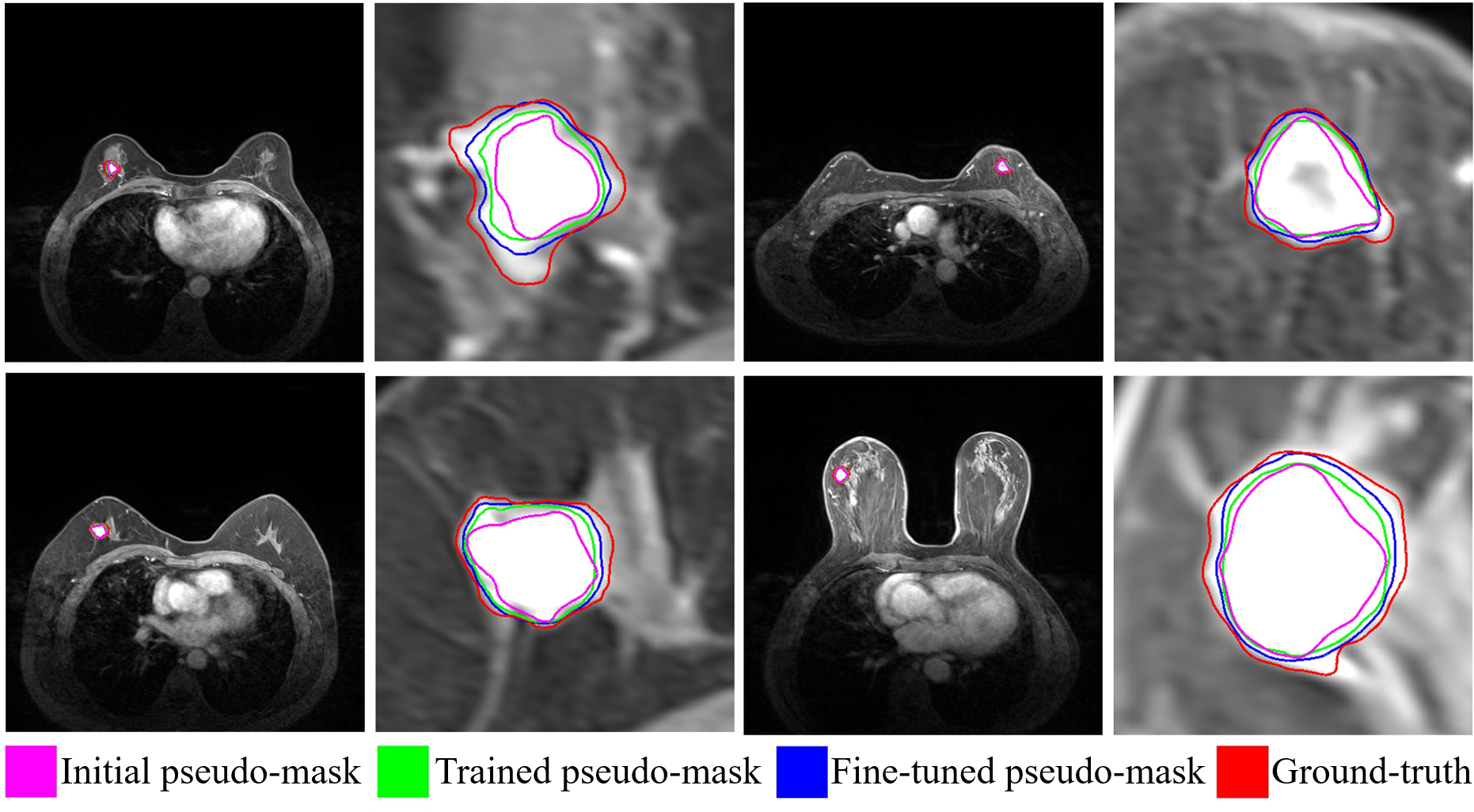}
	\caption{The pseudo-masks (shown as boundaries) at different training stages, including the initial pseudo-mask generated by the random walker (purple), the trained network's output (green), the fine-tuned pseudo-mask using SimPLe (blue) and the ground-truth (red). Note that all images here are the training images.}
	\label{fig3}
\end{figure}

\begin{figure}[t]
	\centering
	\includegraphics[width=1.0\textwidth]{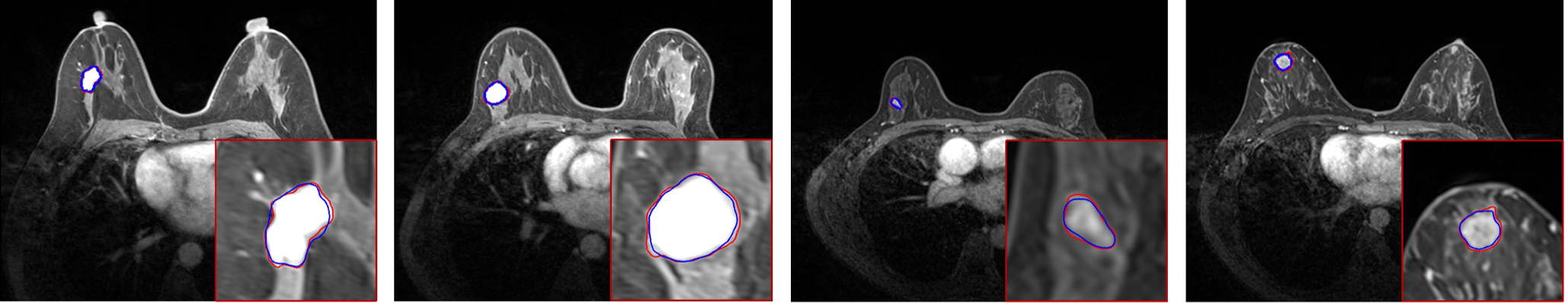}
	\caption{The segmentation visualization in transversal slices. The blue and red contours are the segmented boundaries and the ground-truths, respectively.}
	\label{fig4}
\end{figure}

We first verified the efficacy of our SimPLe in the training stage.
Fig.~\ref{fig3} illustrates the pseudo-masks at different training stages.
It is obvious that our SimPLe effectively updated the pseudo-masks to make them approaching the ground-truths.
Therefore, such fune-tuned pseudo-masks could be used to retrain the network for better performance.

\begin{figure}[t]
\centering
\includegraphics[width=1.0\textwidth]{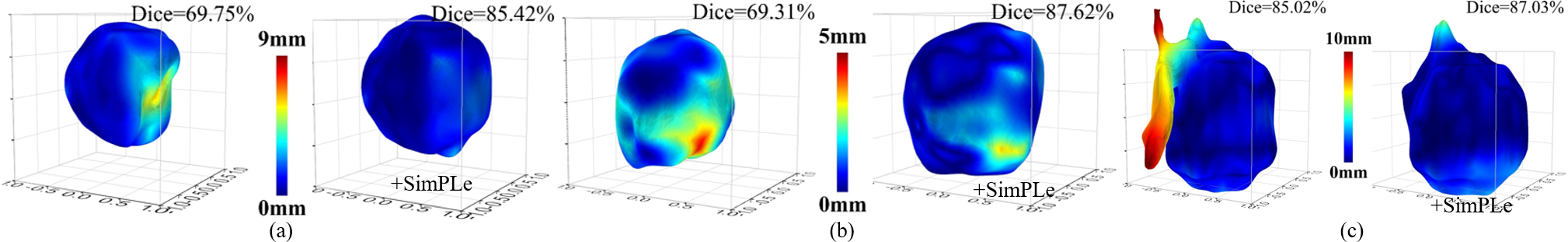}
\caption{
Three cases of 3D visualization of the surface distance between segmented surface and ground-truth. Each case shows the $\mathcal{L}_{pce} + \mathcal{L}_{ctr}$ result and the SimPLe fine-tuned result. The proposed SimPLe consistently enhances the segmentation.}
\label{fig5}
\end{figure}

Fig.~\ref{fig4} visualizes our cancer segmentation results on the testing data.
Table~\ref{tab1} reports the quantitative Dice and Jaccard results of different methods.
We compared our method with an end-to-end approach~\cite{dorent2020scribble} that proposed to optimize network via CRF-regularized loss $\mathcal{L}_{crf}$.
Although our $\mathcal{L}_{ctr}$ supervised method outcompeted $\mathcal{L}_{crf}$~\cite{dorent2020scribble}, the networks trained only using the initial pseudo-masks could not achieve enough high accuracy (Dice values$\textless$70\%).
In contrast, the proposed SimPLe largely boosted the performance of the basically trained networks, by +14.74\% Dice and +15.16\% Jaccard ($v$.$s$. $\mathcal{L}_{crf}$), +11.81\% Dice and +12.65\% Jaccard ($v$.$s$. $\mathcal{L}_{ctr}$).
Furthermore, our method achieved competitive Dice results compared with fully supervision, which again proves the efficacy of the proposed SimPLe strategy.
Fig.~\ref{fig5} visualizes the 3D distance map between the segmented surface and ground-truth. 
It can be observed that our SimPLe consistently enhanced the segmentation.

\begin{table}[t]
\caption{The numerical results of different methods for breast cancer segmentation}
\label{tab1}
\centering
\begin{tabular}{p{4cm}p{3cm}p{3cm}}
    \toprule
    Methods     & Dice {[}\%{]}    & Jaccard {[}\%{]}  \\ 
    \midrule
    $\mathcal{L}_{pce} + \mathcal{L}_{crf}$~\cite{dorent2020scribble}
    & 64.97$\pm$28.66      & 53.83$\pm$27.26 \\
    $\mathcal{L}_{pce} + \mathcal{L}_{ctr}$
    & 69.39$\pm$24.09      & 57.36$\pm$23.31 \\
    \midrule
    $\mathcal{L}_{pce} + \mathcal{L}_{crf}$$+$~SimPLe
    & 79.71$\pm$17.72      & 68.99$\pm$18.84 \\
    $\mathcal{L}_{pce} + \mathcal{L}_{ctr}$$+$~SimPLe
    & 81.20$\pm$13.28      & 70.01$\pm$15.02 \\
    \midrule
    Fully Supervision
    & 81.52$\pm$19.40      & 72.10$\pm$20.45 \\
    \bottomrule
\end{tabular}
\end{table}

\section{Conclusion}
We introduce a simple yet effective weakly-supervised learning method for breast cancer segmentation in DCE-MRI.
The primary attribute is to fully exploit the simple \textit{train} - \textit{fine-tune} - \textit{retrain} process to optimize the segmentation network via only extreme point annotations.
This is achieved by employing a similarity-aware propagation learning (SimPLe) strategy to update the pseudo-masks.
Experimental results demonstrate the efficacy of the proposed SimPLe strategy for weakly-supervised segmentation.

\section*{Acknowledgements}
This work was supported in part by the National Natural Science Foundation of China under Grants 62071305, 61701312 and 81971631,
and in part by the Guangdong Basic and Applied Basic Research Foundation under Grant 2022A1515011241.

%
%
%
\bibliographystyle{splncs04}
\bibliography{reference}

\end{document}